\documentstyle[aps,prb,epsfig,fancyhdr,multicol]{revtex}

\begin{document}



\title{A novel multigrid method for electronic structure calculations}

\author{M.~Heiskanen, T.~Torsti, M.~J.~Puska, and R.~M.~Nieminen}

\address{Laboratory of Physics, Helsinki University of Technology,
P.O, Box 1100, FIN-02015 HUT, FINLAND} 

\date{\today}
\maketitle

\thispagestyle{fancy}

\begin{abstract}
A general real-space multigrid algorithm for the self-consistent solution
of the Kohn-Sham equations appearing in the state-of-the-art electronic-structure 
calculations is described. The most important part
of the method is the multigrid solver for the Schr\"odinger equation.
Our choice is the Rayleigh quotient multigrid method (RQMG), which applies
directly to the
minimization of the Rayleigh quotient on the finest level. Very
coarse correction grids can be used, because there is no need to be
able to represent the states on the coarse levels. The RQMG method is
generalized for the simultaneous solution of all the states of the system
using a penalty functional to keep the states orthogonal. 
The performance of the scheme is demonstrated by applying
it in a few molecular and solid-state systems described by
non-local norm-conserving pseudopotentials.
\end{abstract}
\pacs{71.15.Dx, 31.15.Ew}

\begin{multicols}{2}
\narrowtext

\section{Introduction}
\label{sec:introduction}

One of the goals of computational materials science is to calculate
from first principles the various physical and chemical properties. This
requires the solution of the electronic and ionic structures
of the materials system in question. The density-functional
theory (DFT) makes a huge step towards this goal by casting the 
untractable problem of many interacting electrons
to that of noninteracting particles under the influence of an effective
potential \cite{DFTreviews}. The adiabatic approximation allows one to separate
the ionic degrees of freedom from those of the electrons. However,
in order to apply DFT in practice one has to resort to 
approximations for electron exchange and correlation such as the
local-density approximation (LDA) or the generalized-gradient
approximation (GGA). Moreover, in the case of systems 
consisting of hundreds or more atoms it is still a challenge
to solve numerically efficiently for the ensuing Kohn-Sham 
equations. 

The numerical solution of the Kohn-Sham equations is the concern of our
present work. It deals with real-space (RS) methods, in which the values 
of the different functions are presented using three-dimensional point 
grids, and the partial differential equations are discretized using
finite differences \cite{beckrev,arias}. The RS methods, as suggested 
by the name chosen, are contrasted with the popular plane-wave (PW) 
schemes \cite{paynerev,kresse1}. 
There are several aspects favouring the RS methods over the PW methods. 
Both of the methods are used in the context of pseudopotentials describing the 
electron-ion interactions, but only the RS can easily be
used in all-electron calculations or with hard pseudopotentials
of, {\em i.e.} first-row or transition metal atoms, because the RS 
grid can be refined in a natural way in the ion core regions
(composite grids \cite{bai,bylaska,fattebert2}, adaptive coordinates 
\cite{gygi,kaxiras,waghmare}). 
Systems, such as surfaces, containing different length scales
are more economically described in the RS than in the PW
scheme because one needs not  waste many grid points in
the vacuum regions to describe the slowly varying tails of
wave functions. In the RS methods periodic
boundary conditions are not necessary. This leads to ease and 
accuracy in describing charged atomic clusters in contrast to
PW methods requiring an artificial neutralizing background charge.
Besides the above ``physical arguments'' there are also 
methodological and computational aspects favouring the RS
methods. The RS methods allow a systematic convergence
control by increasing the grid (or basis function) density.
(The PW methods do also so by adjusting the cutoff energy of
the plane wave expansions.) The so-called ``order-N'' methods
\cite{goedecker}, the computational cost of which scales linearly 
with the number of electrons, require localized real-space wave 
functions leading naturally to the employment of RS methods
\cite{fattebert3}. The discretizations in the real-space grid can be 
made local and therefore parallelization can effectively use data
decomposition in which different real-space regions are 
handled with different processing units and the communications
between processing units will be mainly short-ranged 
\cite{briggs2}.

More specifically, our choice for the numerical method is a
multigrid scheme\cite{brandt1,beckrev}. Several approaches employing 
the multigrid idea within electronic structure calculations have 
appeared during recent years 
\cite{briggs1,briggs2,ancilotto,fattebert2,wang1}. 
The main idea of multigrid methods is that
they avoid the critical slowing-down (CSD) phenomenon occuring when
a partial differential equation discretized on a real space grid 
is solved with a simple relaxation method such as the Gauss-Seidel
method. The discretization operators typically use information from 
a rather localized region of the grid at a time. Therefore the 
high frequency error of the length scale of the grid spacing
is reduced very rapidly in the relaxation. However, 
once the high frequency error has effectively been removed, the
very slow convergence of the low frequency components dominates
the overall error reduction rate~\cite{brandt1}, {\em i.e.} CSD occurs.
In multigrid methods one stops the relaxation on a given
(fine) grid before CSD sets in and transfers the equation to a
coarser grid (the so-called restriction operation) where the low-frequency 
components can be solved more efficiently. On the coarsest 
grid the problem is solved exactly or as accurately as possible, 
after which one interpolates (the so-called prolongation operation) 
the correction to finer grids, performing simultaneously relaxations 
in order to remove the high-frequency errors introduced in the 
interpolation.

The solution of the Poisson equation by multigrid methods is 
straightworward \cite{brandt1}. This is because the error (or the 
correction needed) also obeys a Poisson equation and thus will be a 
smooth function to be presented and solved on the repeatedly coarser grids 
optimal to handle the lower frequencies. The solution
of an eigenvalue problem, such as the Schr\"odinger equation,
is a much more complicated task than that of the Poisson equation.
The problem is no more linear because both the eigenfunction and the
eigenvalue have to be solved simultaneously. Then the error 
no longer obeys the same equation as the solution.  
Also one has to solve for several eigenpairs (eigenvalues and corresponding 
eigenvectors). Moreover, the existence of both negative and positive 
eigenvalues makes the problem indefinite. This implies severe difficulties 
for many simple iterative methods which converge  only in the case of
a positive definite iteration matrix.
In particular, it can easily be shown that using
Gauss-Seidel relaxation for the Schr\"odinger equation the high frequency
components typically converge as in the case of the Poisson equation,
but the low frequency components may diverge, although the divergence
may be slow~\cite{grinstein}. More complicated methods
such as Kaczmarz relaxation are guaranteed to converge, but may have
clearly inferior high frequency reduction rates, which are essential
for the overall speed of multigrid methods. Other possible 
convergent methods include GMRES~\cite{saad} 
which is considerably more complex than Gauss-Seidel relaxation.


A standard recipe for dealing with eigenproblems with multigrids
is the full-approximation-storage (FAS) method originally described by 
Brandt~\cite{brandt2}. In FAS one solves for the entire problem on the
coarse grids also and ends up in solving for a properly 
modified problem so that its solution can
be used in correcting the fine grid solution. The FAS method may not
be very straighworward to implement for the Schr\"odinger equation.
It is also difficult to present some actual potential on the
coarse levels accurately enough.
However, some succesful applications of FAS have appeared in the 
context of electronic structure calculations by Beck {\em et al.}
\cite{beck1,wang1} and advanced strategies for FAS have been 
proposed~\cite{costiner}.

Briggs {\em et al.} \cite{briggs1,briggs2} employ a multigrid method 
in electronic structure calculations by 
linearizing the Schr\"odinger problem and presenting the potential 
contribution on the coarse levels by an error term (residual) only. 
Then on the coarse levels they solve effectively for the Poisson problem.
Ancilotto {\em et al.} \cite{ancilotto} modified the method by 
Briggs {\em et al.} by shifting to a full multigrid (FMG) scheme and by 
solving on the coarse grids a problem including a local potential term.  
The idea of FMG is to start the smoothing iterations from a
coarse grid. Then the interpolation to a finer grid provides a good
initial guess of the solution. The FMG scheme can accelerate the
convergence remarkably with respect to the (above-described) V-cycle
scheme in which one starts from the finest level. 
Fattebert~\cite{fattebert2} used a multigrid method with a block 
Galerkin inverse iteration (BGII) and GMRES in the relaxations.
In the method, the current approximation 
is kept orthogonal against all the nearby states during the multigrid cycle. 
The inverse iteration converges for a given guess for the energy
eigenvalue towards the nearest eigenvalue. In order to
solve all the desired lowest eigenstates a good guess for the
eigenvalue spectrum is needed in the beginning of iterations, but
thereafter large computational savings may be expected because explicit
orthogonalizations are not needed (at least between well-separated 
states).

A severe problem in the existing multigrid schemes for the Schr\"odinger equation
is often that the coarse grids cannot well approximate the solutions of 
the coarse grid equations themselves. As a consequence the correction from
coarse grids, no matter how accurately the equation is solved, may
be ineffective in correcting the fine grid solution and as a result
the overall process converges slowly. Therefore 
one has to restrict to the use of rather fine grids only and
the convergence speed of the scheme is drastically lowered. 
In those multigrid methods, which use the potential also on the coarse grids 
the size of the coarsest grid has been typically of the order of 31x31x31
\cite{fattebert2,ancilotto}. Using the FAS method coarser grids are
possible at least for systems with a small number of 
eigenstates solved \cite{wang1}. If a large number of eigenstates
have to be solved problems may arise because the coarse grids may
not be able to represent eigenstates with many nodes or the ordering
of the states may change between the successive grids. To bypass these
problems in FAS, rather complicated strategies are needed \cite{costiner}. 

In order to avoid the coarse grid representation problems we 
utilize the so-called Rayleigh Quotient Multigrid 
(RQMG) method introduced by Mandel and McCormick \cite{McCormick}.
In this method the coarse grid relaxation passes
are performed so that the Rayleigh quotient calculated on the {\em fine}
grid will be minimized. In this way there is no requirement for the
solution to be well represented on a coarse grid and the coarse grid
representation problem is avoided. Mandel and McCormick\cite{McCormick}
introduced the method for the solution of the eigenpair corresponding to 
the lowest eigenvalue. We have generalized it to the simultaneous
solution of a desired number of lowest eigenenergy states by developing
a scheme which keeps the eigenstates separated by the use of a
penalty functional, Gramm-Schmidt orthogonalization, and subspace
rotations. Our generalization of the RQMG method
is an attractive alternative for large-scale electronic structure
calculations. 

The Kohn-Sham equations have to be solved self-consistently, {\em i.e.}
the wave functions solved from the single-particle equation 
determine via the density (solution of the Poisson equation 
and the calculation of the exchange-correlation potential)
the effective potential for which they
should again be solved. To approach this self-consistency
requires an optimized strategy so that numerical accuracy of the wave 
functions and the potential increase in balance, enabling the most 
efficient convergence \cite{wang1}. In order to avoid the divergence
of the self-consistency iterations, the mixing of the input and
output solutions is needed. For this feedback procedure sophisticated
schemes \cite{kresse2} and control strategies \cite{waghmare} have
been presented. 

The outline of the present paper is as follows. In Section II we
represent shortly the most important ideas of the density functional
theory. Section III is devoted for numerical methods, the most important
of which is the Schr\"odinger equation solver developed. Also the strategy
for the self-consistency iterations is discussed. In Section IV we 
demonstrate by the help of a couple of examples the performance of our 
scheme in calculating the electronic structures of small molecules and 
solid-state systems described by pseudopotentials. Section V summarises
the work and gives outlines for the future developments.

\section{The Kohn-Sham scheme}

In the Kohn-Sham method for electronic structure calculations
one solves for a set of equations self-consistently \cite{DFTreviews}.
In the following, we present the equations in the spin-compensated
form. In practice, we have made the straightforward generalization
using the spin-density functional theory. The set of equations
reads as (atomic units with $\hbar = m_e = e = 1$ are used):
\begin{equation}
\label{kohnshameq}
 \left(-\frac{1}{2}\nabla^2 +
        V_{\rm eff}({\bf r})
       \right) \Psi_i = \epsilon_i \Psi_i,
\end{equation}
\begin{equation}
\label{densitydef}
  n({\bf r}) = \sum\limits_i^N |\Psi_i({\bf r})|^2,
\end{equation}
\begin{equation}
\label{veffdef}
V_{\rm eff}({\bf r}) = V_{\rm ion}({\bf r}) +
 V_{\rm H}({\bf r}) + V_{\rm XC}({\bf r}),
\end{equation}
\begin{equation}
\label{hartreedef}
  V_{\rm H}({\bf r}) = \int \frac{n({\bf r}')}
         {|{\bf r}-{\bf r}'|}d{\bf r}',
\end{equation}
\begin{equation}
\label{vxcdef}
  V_{\rm XC}({\bf r}) = \frac{\delta E_{\rm XC}[n({\bf r})]}
      {\delta n({\bf r})}.
\end{equation}
The first equation (\ref{kohnshameq}) is a Schr\"odinger equation for 
non-interacting particles in an effective potential 
$V_{\rm eff}({\bf r})$. For finite systems the wave functions
are required to vanish at the boundaries of the computation volume.
In the case of infinite periodic systems the complex wave functions 
have to obey the Bloch theorem at the cell boundaries.
The electron density $n({\bf r})$ is
obtained from a sum over the $N$ occupied states. The effective potential
consists of an external potential $V_{\rm ion}({\bf r})$
due to ions (or nuclei in all-electron calculations), the
Hartree potential $V_{\rm H}({\bf r})$ calculated from the electron 
density distribution, and the exchange-correlation potential
$V_{\rm XC}({\bf r})$. In the examples of the present work we use 
the norm-conserving non-local pseudopotentials for the electron-ion
interactions and the local-density approximation (LDA) for the 
exchange-correlation energy
\begin{equation}
E_{\rm XC} [n({\bf r})] = \int \epsilon_{\rm XC}(n({\bf r})) n({\bf r})
                 d{\bf r}, 
\end{equation}
and for the exchange-correlation potential
\begin{equation}
V_{\rm XC}({\bf r}) = \epsilon_{\rm XC}(n({\bf r})) + n({\bf r})
              \frac{d \epsilon_{\rm XC} }{d n} _{\vert n = n({\bf r})}.
\end{equation}

The Hartree potential is solved from the Poisson equation
\begin{equation}
  \nabla^2 V_{\rm H}({\bf r}) = -4\pi n({\bf r}).
\end{equation}
In practice, the electron density $n({\bf r})$ is substituted
by the total charge density $\rho({\bf r})$, which includes
the positive ionic (nuclear) charge neutralizing the system. 
In the case of finite systems, Dirichlet boundary conditions are used with 
the Coulomb potential values calculated using a multipole expansion.
For periodic systems we fix the average Coulomb potential to zero
and allow the periodic boundary conditions to result in the corresponding
converged potential. 

The self-consistent solution of the above Kohn-Sham equations leads
to the ground state electronic structure minimizing the total energy
\label{kohnshamenergy}
\begin{eqnarray}
  E_{\rm tot}  & = &
  \sum\limits_{i}^{ }\int \Psi_i^*({\bf r})\left(-\frac{1}{2}\nabla^2\right)
  \Psi_i({\bf r})d{\bf r}
    +\frac{1}{2}\int V_{\rm H}({\bf r}) n({\bf r}) d{\bf r} \nonumber \\ 
 &  + & \int V_{\rm ion}({\bf r}) n({\bf r}) d{\bf r}
    + E_{\rm XC} + E_{\rm ion-ion} \ ,
\end{eqnarray}
where $E_{\rm ion-ion}$ is the repulsive interaction between the ions
(nuclei) of the system. Instead of the self-consistency iterations the
solution of the Kohn-Sham problem can be found by minimizing directly
the total energy with respect to the wave function parameters, 
{\em e.g.} plane-wave coefficients \cite{paynerev}. However,
Kresse and Furthm\"uller \cite{kresse1,kresse2} have found this scheme 
less efficient than the self-consistency iterations.

\section{Numerical methods}
\label{sec:methods}

\subsection{Schr\"odinger equation solver}

In our real space method we start
from an initial guess for the effective potential and initial
wave functions generated by random numbers in grid points.
The wave functions and the Hartree potential are updated 
alternatingly towards self-consistency. The solution of the
Poisson equation is a standard task for the multigrid scheme.
If a reasonable guess for the Coulomb potential is not available, 
the FMG method will produce the solution starting
from random numbers and requiring the work which scales 
linearly as a function of the size of the system ($O(N)$).
During the Kohn-Sham iterations one can start from the 
present approximation of Coulomb potential and update it
with respect to the new charge density by performing only a few
V-cycles.

The solution of the wave functions is a much more complicated
task than that of the Poisson equation
because one has to solve an eigenvalue problem which
in the state-of-the-art electronic structure calculations means
the determination of several hundreds of eigenpairs.
For this purpose we have developed a scheme based on 
RQMG method introduced by Mandel and McCormick\cite{McCormick} 
for the solution of the eigenpair corresponding to the lowest eigenvalue.
We begin by reviewing the basic principles of RQMG. This
is most easily done in the framework of the so-called
coordinate relaxation method. Thereafter we go through the
modifications made in order to simultaneously solve for
several eigenpairs.

Coordinate relaxation is a method of solving the discretized
eigenproblem
\begin{equation}
  H u = \lambda B u
\end{equation}
by minimizing the Rayleigh quotient
\begin{equation}
\label{Ray}
  \frac{\langle u\arrowvert H\arrowvert u\rangle}
       {\langle u\arrowvert B\arrowvert u\rangle}.
\end{equation}
Above, $H$ and $B$ are matrix operators chosen so that the Schr\"odinger
equation discretized on a real-space point grid with spacing $h$
is satisfied to a chosen order $O(h^n)$. 
In Eq. (\ref{Ray}) $u$ is a vector containing
the wave function values at the grid points. In the relaxation 
method, the current estimate $u$ is replaced by itself plus a 
multiple of some search vector $d$
\begin{equation}
 \label{rqsubsteq1}
 u' = u + \alpha d,
\end{equation}
and $\alpha$ is chosen to minimize the Rayleigh quotient. This leads
to a simple quadratic equation for $\alpha$. (Find the minimum of the
expression (\ref{rqmgeq}) below with respect to $\alpha$. In the case
of a complex wave function one has to solve for the real and imaginary
parts of $\alpha$ from a coupled pair of quadratic equations.) Moreover,
the search vector $d$ is simply chosen to be unity in one grid point
and to vanish in all other points. A complete coordinate relaxation 
pass is then obtained by performing the minimization at each point 
in turn and these passes can be repeated until the lowest state is 
found with desired accuracy.

Naturally, also the coordinate relaxation suffers from CSD
because of the use of local information only in
updating $u$ in a certain point. In order to avoid it one applies the 
multigrid idea. In the multigrid scheme by Mandel and 
McCormick\cite{McCormick} 
the crucial point is that {\em coarse} grid coordinate relaxation passes
are performed so that the Rayleigh quotient calculated on the {\em fine}
grid will be minimized. In this way there is no requirement for the
solution to be well represented on a coarse grid. In practice, a
coarse grid search substitutes the fine grid solution by
\begin{equation}
\label{rqmgchgeq}
 u_f' = u_f + \alpha I_c^f d_c,
\end{equation}
where the subscripts $f$ and $c$ stand for the fine and coarse
grids, respectively, and $I_c^f$ a prolongation operator interpolating
the coarse grid vector to the fine grid. The Rayleigh quotient to
be minimized is then 
\begin{eqnarray}
\label{rqmgeq}
  &  \frac{\langle u_f + \alpha I_c^f d_c \arrowvert H_f \arrowvert
         u_f + \alpha I_c^f d_c \rangle}
       {\langle u_f + \alpha I_c^f d_c \arrowvert B_f \arrowvert
         u_f + \alpha I_c^f d_c \rangle} = \qquad \qquad \qquad \qquad \nonumber \\
&\qquad \qquad \qquad      \frac{ \langle u_f \arrowvert H_f u_f \rangle
         + 2\alpha \langle I_f^c H_f u_f \arrowvert d_c \rangle
         + \alpha^2 \langle d_c \arrowvert H_c d_c \rangle
       }
       { \langle u_f \arrowvert B_f u_f \rangle
         + 2\alpha \langle I_f^c B_f u_f \arrowvert d_c \rangle
         + \alpha^2 \langle d_c \arrowvert B_c d_c \rangle
       }.
\end{eqnarray}
The second form is obtained by relating the coarse grid operators,
$H_c$ and $B_c$, with the fine grid ones, $H_f$ and $B_f$, by the
Galerkin condition
\begin{equation}
\label{galerkincond}
 \begin{array}{ll}
   H_c & = I_f^c H_f I_c^f \\
   B_c & = I_f^c B_f I_c^f,
 \end{array}
\end{equation}
and the restriction operator $I_f^c$ has to be the transpose of the 
prolongation operator
\begin{equation}
 I_f^c = \left(I_c^f\right)^T.
\end{equation}
The key point to note is that when $H_f u_f$ and $B_f u_f$ are provided 
from the fine grid to the coarse grid, the remaining
integrals can be calculated on the coarse grid itself. Thus one really
applies coordinate relaxation on the coarse grids to minimize the
\textit{fine level} Rayleigh quotient. This is a major departure from
the earlier methods, which to some extent rely on the ability to
represent the solution of some coarse grid equation on the coarse grid
itself. Here, on the other hand, one can calculate the \textit{exact}
change in the Rayleigh quotient due to \textit{any} coarse grid
change, no matter how coarse the grid itself is. There is no equation
whose solution would have to be representable.

Thus, in the Rayleigh quotient minimization multigrid (RQMG) algorithm
the coordinate relaxation passes on each level keep track of the 
integrals in Eq.~(\ref{rqmgeq}). Actually, on the finest
level we use Gauss-Seidel relaxation, which very effectively
smoothens the errors of the wavelength corresponding to the grid spacing.
When calculating several eigenpairs Gauss-Seidel relaxation may also 
work as a {\em residual minimization} method. The idea is that the 
coarse grid-iterations with Gramm-Schmidt orthogonalization can provide 
the separation of the eigenstates so well that the subsequent finest
level relaxations converge to the correct (nearest) eigenstates
without orhogonalization. This requires that the effect of the coarse-level
smoothings on the low-frequency components of the solutions overcomes 
the possible divergence tendency of these components caused by 
the Gauss-Seidel relaxation on the finest level.

Moreover, we discretize the original equation separately on each grid
(discretization coarse grid approximation (DCA))
instead of using the Galerkin conditions of Eq. (\ref{galerkincond})
This may in principle decrease the convergence rate and force a limit
to the coarsest possible grid in order to avoid instability or 
divergence. However, we have observed this DCA implementation of RQMG 
to be quite stable and efficient. To avoid possible coarse-level 
instabilities occuring especially during the first few iteration cycles
we may recalculate the Rayleigh quotient whenever coarse grid corrections 
are interpolated to a finer grid.
Later when approaching the convergence the recalculation can be omitted.

For the matrix operators $H$ and $B$ we have used either high-order
($O(h^4)$ or higher) Mehrstellen or central difference stencils (CDS)
\cite{briggs2,fattebert2}. The use of high-order stencils reduces
remarkably the density of grid points needed. The benefit of the
Mehrstellen scheme is that more local information is used. The scheme 
leads to controlled accuracy and convergence properties and 
to more isotropic smoothing of the error in comparison with the use of CDS's. 
The local nature enables also a more efficient parallel coding. As the
prolongation operator $I_c^f$ we usually use trilinear interpolation and as the
restriction operator $I_f^c$ its transpose, the so-called full-weighting 
operator, in which the coarse-grid values are chosen to be
the averaged values of the surrounding fine grid points. The integrations
are performed by the trapezoidal rule.

Next we consider the generalization of the RQMG method to the simultaneous
solution of several ($N$) mutually orthogonal eigenpairs. The separation
of the different states is divided into two or three subtasks. First,
in order to make the coarse grid relaxations converge towards the 
desired state we apply a penalty functional scheme.
Given the $k$ lowest eigenfunctions, the next lowest, $(k+1)$'th  state 
is searched for by minimizing the functional
\begin{equation}
\label{rqmgneq}
 \frac{\langle u_{k+1}\arrowvert H\arrowvert u_{k+1}\rangle}
      {\langle u_{k+1}\arrowvert B\arrowvert u_{k+1}\rangle}
 + \sum\limits_{i=1}^{k}
    q_i \frac{\langle u_i | u_{k+1}\rangle^2}
             {\langle u_i | u_i\rangle \cdot
              \langle u_{k+1} | u_{k+1}\rangle}.
\end{equation}
The overlap integral in
the penalty term is squared to make the penalty positive definite. 
The denominator is required to make the functional
independent of the norms of $u_i, \, i=1\ldots k+1$.
The minimization of this functional is equivalent to imposing
the orthonormality constraints against the lower $k$ states,
when $q_i \rightarrow \infty$. By increasing the shifts $q_i$
any desired accuracy can be obtained, but in order to obtain
a computationally efficient algorithm a reasonable finite value should
be used, for example
\begin{equation}
  q_i = (\lambda_{k+1}-\lambda_i) + {\rm Q},
\end{equation}
where $Q$ is a sufficiently large positive constant. In our test 
calculations $Q$ is of the order of $Q=0.5\ldots 2$ Ha.

We minimize the expression~(\ref{rqmgneq}) \textit{simultaneously} 
for all $N$ states. This simplifies the algorithm and
enables a future parallelization over the eigenstates.  
Thus the current approximations are used
for $u_i, \, i=1\ldots k$. Moreover, changes in the $u_i$ during a
given relaxation sweep are not used to update the penalty term in
Eq. ~(\ref{rqmgneq}). This is sufficient, when the states are
always ordered in the same way, in the order of increasing 
eigenvalue. In order to reduce computations, the $B$-innerproduct is 
actually used in calculating the penalty term integrals because the 
values of $Bu$ are readily available from the finer level.
The substitution (\ref{rqmgchgeq}) is introduced in the functional
~(\ref{rqmgneq}) and the minimization with respect to $\alpha$
leads again to a quadratic equation. This time the coefficients
contain terms due to the penalty part. 

On the finest level, we do not apply the minimization of the
penalty functional. The ideal situation would be if a residual minimization
method, such as the Gauss-Seidel method, would keep the states
calculated on the coarse levels separated. We found out in practical
calculations that this is not true at least when the states are
far from convergence. Therefore we have developed for 
the finest level a scheme which by employing Gramm-Schmidt 
orthogonalization and subspace rotation keeps the eigenstates orthogonal.
The subspace rotation is a method to find the most
optimally separated eigenvectors from the approximative ones.
The major steps of the rotation are:

(i) Calculation of the Hamiltonian matrix elements between the current
states:
\begin{equation}
\label{rot1}
{\bar H}_{i,j} =  \langle u_i|B^{-1}H|u_j\rangle. 
\end{equation}

(ii) Calculation of the overlap matrix:
\begin{equation}
\label{rot2}
{\bar S}_{i,j} =  \langle u_i|u_j\rangle.
\end{equation}
The use of matrix elements of Eqs. (\ref{rot1}) and (\ref{rot2}) leads to 
eigenvectors orthogonal in the desired Euclidian sense ($I$-orthogonal) 
and not in the sense of the $B$-innerproduct.

(iii) Diagonalization to find the optimal eigenvectors 
($u'_k = \sum_j {\bar A}_{k,j} u_j$) and corresponding eigenvalues 
($\lambda_k$):
\begin{equation}
\sum_j {\bar H}_{i,j} {\bar A}_{j,k} 
= \lambda_k \sum_j {\bar S}_{i,j} {\bar A}_{j,k}.
\end{equation}
In practice, we apply the approximation
\begin{equation}
\langle u_i|B^{-1}H|u_j\rangle \approx {\langle u_i|u_j\rangle}
    \frac{\langle u_i|H u_j\rangle}{\langle u_i|B u_j\rangle}.
\end{equation}

The Gramm-Schmidt orthogonalization and the subspace rotations
are organized so that the space of the eigenvectors is first divided to
small clusters corresponding to close eigenvalues. The Gramm-Schmidt 
orthogonalization is then performed for each cluster at a time so that 
its eigenvectors become orthogonal against the eigenvectors of the
clusters of lower eigenvalues. Then a subspace rotation is performed 
within the states belonging to the present cluster. The division
to clusters reduces remarkably the cost of the subspace rotation.
This is because the cost is proportional to $O(N^3$), where $N$ is 
the number of states rotated. Moreover, the subspace rotation 
requires the calculation of matrix elements which are more complicated
than those for the simple Gramm-Schmidt orthogonalization.

According to our test calculations this subspace rotation scheme
leads quite effectively to $I$-orthogonal eigenstates. This is
seen as a convergence of the eigenvalue problem within the numerical
accuracy, {\em i.e.} the residuals of different eigenstates vanish.
In order to prove this one has, in practice, to introduce potential 
shifts which reduce the number of significant digits in the eigenvalues
so that the numerical accuracy of the eigenvalue does not prevent to
reach the numerical accuracy of the wavefunction, {\em i.e.} the
vanishing residual. The error in the eigenvalue scales as the square
of the residual. When applying the subspace rotation it is 
important to complete the highest eigenvalue cluster; otherwise the rotation 
may become inefficient.

The orthogonalization
needed scales as $O(N^3)$. For small systems of several tens of
eigenpairs this is not yet a problem. The algorithm is effective
and the number of fine grid orthogonalizations remains quite
plausible, for example, in comparison with the conjugate gradient
search of eigenpairs employing only the finest grid \cite{seitsonen}.
But for larger systems with
hundreds of states it will be the bottleneck. One solution could
be to rely on the finest level only on a residual minimization
method when the initial stages of the iteration process have been
performed and the solution is clearly on a stable track towards
convergence. 

\subsection{Strategy for self-consistency iterations}
\label{sec:strategy}

The Kohn-Sham problem has to be solved self-consistently. This means
that an optimal strategy is needed so that computing time is not wasted in the
beginning of the self-consistency iterations to obtain unnecessarily accurate 
wave functions, because these will change during the later iterations
due to the changes in the potential. Updating the potential,
including the solution of the Poisson equation, is a much less
time-consuming task than the update of all the wavefunctions. Therefore
the potential update can be performed frequently \cite{wang1}.

\begin{figure}
\resizebox{\columnwidth}{!}{\includegraphics{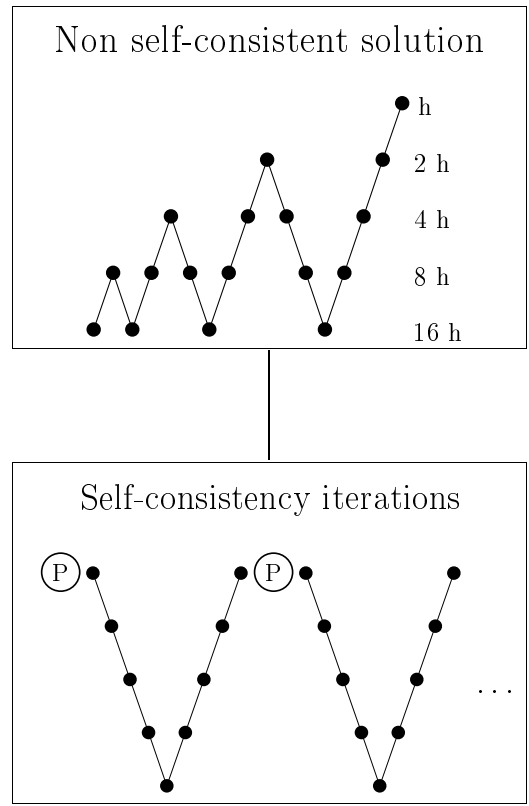}}
\caption{Strategy of self-consistency iterations. First, the wavefunctions are solved nonselfconsistently
using the full multigrid method in the initial 
potential corresponding to the superposition of pseudoatoms. 
Then the effective potential is updated (this is denoted by P in the figure). 
The potential update amounts to calculation of the new electron density, 
the solution of the Poisson equation and calculation of the new exchange correlation 
potential. Next the wave-functions are updated by one V-cycle. These two steps are repeated
until self-consistency has been reached.}
\label{fig:strategy}
\end{figure}

The examples of this paper are small-molecule and
bulk-solid systems described by pseudopotentials. The strategy used is
schematically presented in Fig. \ref{fig:strategy}. Similar strategies can
certainly be applied in other kind of Kohn-Sham calculations, for example
in those employing all-electron or jellium-type models.
In the examples of this work the initial electron 
density is the superposition of the pseudoatom densities centered around given 
nuclear positions. From the superposition we calculate the initial effective 
potential, where the wavefunctions are solved accurately enough using the 
full-multigrid method. The FMG process is started from random numbers
for the wavefunctions on the coarsest level. The accuracy of the 
wavefunctions is controlled by calculating the norms of the 
residuals of the eigenstates and it is finally improved by adding more
V-cycles starting from the finest level. A certain accuracy is needed in order to
initiate self-consistency iterations which converge without large 
density oscillations. Then the  new electron density and the ensuing effective
potential are calculated. The new potential is not directly fed into
the next iteration but it is mixed in this place,
as well as later between the self-consistency iterations,
with the input potential of the iteration. 
We monitor the accuracy
of the wave functions by calculating their residuals and require that
the accuracy has improved from the previous iteration. Usually one V-cycle
is sufficient for this, because the changes in the potential are small.

An important point is also to find a proper balance with respect to the
pre- and postsmoothening sweeps on the different grid levels. Typically, on the
finest level we made two pre- and postsmoothening sweeps whereas on the
coarser grids their number is four. Actually, this means that on the
finest level four successive sweeps are done if the potential is not
updated. A potential update is always preceded and followed by two
immediate  smoothening sweeps.

\section{Tests}
\label{sec:results}

We test the performance of our scheme by calculating the self-consistent
electronic structures of a CO$_2$ molecule as well as that of perfect
bulk Si lattice with a supercell of 64 Si atoms. The former system is 
an example of the employment of Dirichlet boundary conditions and the use of
``hard'' pseudopotentials whereas the latter system represents the
use of periodic boundary conditions and a supercell size typical in
electronic structure calculations for point defects in solids.

The ions are described by pseudopotentials of the Kleinman-Bylander 
form \cite{pseudo},
\begin{eqnarray}	
V_{ion} ({\bf r})& = & \sum_a V_{ion,loc}(|{\bf r}_a|)  	\nonumber \\
	&+& \sum_{a,n,lm}\frac{1}{\bigl< \Delta V_{lm}^a \bigr >} 
	\bigl | \Delta V_{ion,l}(r_a) u_{lm}({\bf r}_a) \bigr > \nonumber \\
	& \times & \bigl < \Delta V_{ion,l}(r'_a) u_{lm}({\bf r'}_a) \bigr | \ ,
\end{eqnarray}
where $\bigl<\Delta V_{lm}^a\bigr>$ is a normalization factor,
\begin{equation}
\bigl<\Delta V_{lm}^a\bigr> = \int u_{lm}({\bf r}_a) 
	\Delta V_{ion,l}(r_a) u_{lm}({\bf r}_a) d^3r,
\end{equation}
and ${\bf r}_a = {\bf r - R}_a$, $u_{lm}$ are the atomic pseudopotential
wave functions of angular and azimuthal momentum quantum numbers $(l,m)$,
from which the $l$-dependent ionic pseudopotentials $V_{ion,l}(r)$ are 
generated using the Troullier-Martins scheme \cite{troullier-martins}. 
The ion core is assumed to be spherically symmetric. 
$\Delta V_{ion,l}(r) = V_{ion,l}(r) - V_{ion,loc}(r)$	 is the difference between the
$l$-component of the ionic pseudopotential and the local ionic potential.
We have chosen the s-component of the pseudopotential as the local component.

Because the functions $u_{lm}({\bf r})$ are short-ranged, operating on the 
wave-function by the nonlocal parts of the pseudopotential is in practice 
a multiplication by a sparse matrix. The numerical work required 
to compute this scales as the square of the number of atoms in the system, 
whereas in the conventional reciprocal-space formulation the work scales as 
the cube of the system size. The advantage of implementing the nonlocal 
pseudopotentials in real space has been noted also in the context of 
plane-wave methods \cite{king-smith}.

In the previous multigrid implementations of the pseudopotential method 
\cite{ancilotto,briggs2}, the nonlocal parts have
only been employed on the finest grid.  It is, however, straightforward 
to implement them also on the coarse levels, and we have found that this 
may increase the convergence rate and stability of the method. 

The CO$_2$ molecule is placed diagonically in the center of a cubic
computation volume of the size of (12.6 a$_{\rm 0})^3$. Experimental
bond lengths are used. Dirichlet boundary 
conditions are used so that the potential values outside the cube are
obtained from a multipole expansion of the charge density. The point mesh 
used is 63$^3$, giving the grid spacing $h$ = 0.20 a$_{\rm 0}$.
The Mehrstellen discretization by Briggs {\em et al.} \cite{briggs2} is used.


In this calculation we used a mixing scheme, where the new effective potential
$V_{in}^{i+1}$ is obtained from the input and output potentials according to
\begin{equation}
V_{in}^{i+1} = (1-\kappa) V_{in}^i + \kappa V_{out}^i.
\label{mixing}
\end{equation}
\begin{figure}
\resizebox{0.85\columnwidth}{!}{\includegraphics{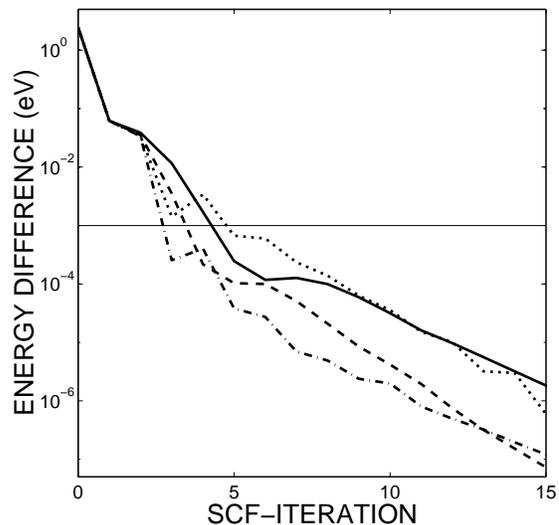}}
\caption{Convergence of the total energy for the CO$_2$~-molecule using
direct mixing with different values of the feedback parameter $\kappa$;
$\kappa = 0.4$ (solid line), $\kappa = 0.5$ (dashed line), 
$\kappa = 0.6$ (dashdotted line) and $\kappa = 0.7$ (dotted line).
A horizontal line has been added to indicate the chemical accuracy of 1 meV.}
\label{CO2}
\end{figure}
The convergence of the self-consistency iterations employing the strategy
described above (Fig. \ref{fig:strategy}) is shown in Fig. \ref{CO2}.
The deviation of the total energy from the converged value is given
as a function of self-consistency iteration steps performed. 
The zeroth iteration is a full-multigrid solution for the wavefunctions 
in the initial potential. Two V-cycles are performed on the finest level 
at this point. The effective potential obtained from the output electron 
density was mixed with the initial potential using the feedback $\kappa = 0.4$.
Next, at iteration one, the wave-functions are relaxed in this new potential
using one V-cycle. From this point on, the four curves in the figure give the
convergence with different values of the feedback parameter $\kappa$.
One V-cycle per one self-consistency iteration step
is done. A wide range of values for $\kappa$ gives satisfactory convergence
indicating a robust behaviour for the scheme. The accuracy of 1 meV,
which is sufficient in practical calculations, is reached after three or four
V-cycles. The implementation of the non-local parts of the pseudopotential 
on the coarse levels is found to speed up the convergence especially in this
region. From Fig. \ref{CO2} we obtain an average convergence rate of 
approximately one decade per self-consistency iteration. This is of the same 
order as those reported by Wang and Beck \cite{wang1} in their FAS scheme 
or by Kresse and Furthm\"uller \cite{kresse2} in their pseudopotential scheme 
employing self-consistency iterations. The convergence rate of one decade per
self-consistency iteration is better than that obtained by Ancilotto
{\em et al.} \cite{ancilotto} in the FMG scheme and much better than
the rate reached in the linearized multigrid scheme by Briggs {\em et al.} 
\cite{briggs2}.

\begin{figure}
\resizebox{0.9\columnwidth}{!}{\includegraphics{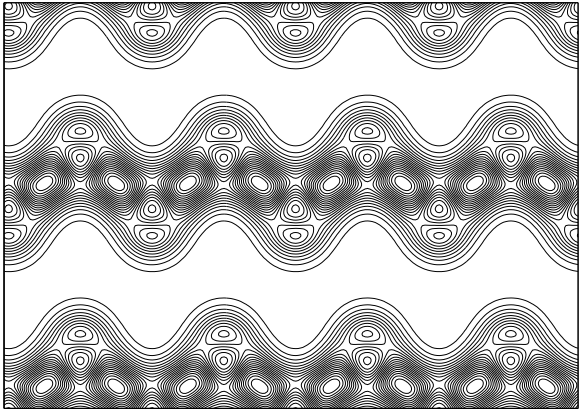}}
\caption{Valence electron density in the (110)-plane obtained in the 
$\Gamma$-point calculation for the 64-atom supercell of bulk Si. 
The area of the figure corresponds to the extent of the supercell.}
\label{Si110}
\end{figure}

We have solved for the electronic structure of perfect Si lattice described 
by a supercell of 64 Si ions. The lattice constant of 20.38 a$_{\rm 0}$
used is the equilibrium value obtained in a plane-wave calculation,
with which we have compared our real-space results. The first Brillouin zone 
is sampled in this test using the $\Gamma$-point only. The point mesh 
used for the wave-functions is 64$^3$, giving the grid spacing 
$h$ = 0.32 a$_{\rm 0}$. For the densities and potentials we use a finer 
grid of 128$^3$ points. The other numerical parameters and the
iteration strategy are the same as in the CO$_2$ test.
The resulting valence electron density on the (110)-plane is given 
in Fig. \ref{Si110}. The area of the figure corresponds to the extent   
of the supercell. One notes that exactly the same features are
reproduced at the equivalent points in different regions of the
supercell. This means that a fully converged result has been 
found. We have compared the results of our real-space code to
those obtained using the plane-wave method. The energy cutoff,
18 Ry, of the plane-wave expansion was chosen so that it results in
a real-space point mesh of 64$^3$, {\em i.e.} it is the same as
in our real-space calculation. The widths of the valence band
and band gaps obtained by the two methods agree with an accuracy
of 3 meV. In the case of degenerate eigenstates the 
real-space code results in degenerate eigenenergies with an accuracy
better than 1 meV. The convergence towards to the self-consistent
solution occurs similarly as for the CO$_2$ molecule in Fig. \ref{CO2}.
Thus, the convergence process seems to be independent of the size
of the system.

\section{Summary and outlook}
\label{sec:conclusions}

In this work we have generalized the RQMG method introduced by Mandel and
McCormick \cite{McCormick} for  the simultaneous solution of a desired 
number of lowest eigenenergy states. This approach can be viewed as belonging 
to a third group of multigrid methods, in addition to FAS and the techniques where the
eigenproblem is linearized ({\it e.g. inverse iteration}. 
In principle, one can use arbitrarily coarse grids in RQMG, whereas in
the other multigrid methods one has to be able to represent all the
states on the coarsest grid.

We have demonstrated the feasibility of the method by electronic
structure calculations for the CO$_2$ molecule and bulk Si described
by pseudopotentials. 
Our strategy for the self-consistent solution consists of a 
full-multigrid solution for the wave-functions in the initial potential,
and subsequent self-consistency iterations. Less than five V-cycles are
generally sufficient for practically sufficient accuracy. The cpu-times
required for the FMG and SCF steps are roughly equal.  

We have applied the 
method also in two-dimensional problems for quantum dots  employing the 
current-spin-density functional theory, in three-dimensional 
cylindrically symmetric systems, and also for calculation
of positron states in solids.

We believe that our method will eventually compete with the standard 
plane-wave methods for electronic structure calculations. However, some 
straightforward programming is still required. For calculations, where 
the optimization of the ionic structure is necessary, the Hellmann-Feynman
forces will be implemented. In order to remove the spurious dependence 
of the total energy on the position of the atoms with respect to the grid 
points, Fourier-filtering of the pseudopotentials is required. Complex 
wave functions for any k-point are easily implemented, and are already in
use in two-dimensional geometries.

Parallelization over k-points can be done easily. One only needs to 
communicate the electron density and effective potential at the end of 
each V-cycle. During the RQMG V-cycle, the states are all relaxed 
simultaneously and independently of each other. Therefore
parallelization over states is natural and easy to implement. However, 
for larger systems the Gramm-Schmidt orthogonalization becomes very 
inefficient in a state-parellel code.  The most efficient 
and yet straightforward choice is real-space data decomposition, 
where each processor is mapped to a specific region of space.

\acknowledgments

M.H. acknowledges stimulating discussions with Prof.  S. F. McCormick 
and Prof. T. L. Beck. 
T.T. acknowledges financial support by the Vilho, Yrj\"o and Kalle 
V\"ais\"al\"a foundation.
This research has been supported by the Academy of Finland
through its Centre of Excellence Programme (2000 - 2005).


\end{multicols}

\end{document}